\documentclass[namedreferences]{solarphysics}

\usepackage[hyperref,optionalrh]{spr-sola-addons} 
\usepackage{graphicx}        
\usepackage{amssymb}        
\usepackage{color}           
\usepackage{breakurl}        
\chardef\us=`\_

\begin{document}

\begin{article}
\begin{opening}

\title{A search for counterparts of quiet sun radio transients in extreme ultraviolet data}

\author[addressref={aff1},corref,email={surajit@ncra.tifr.res.in}]{\inits{S.M.}\fnm{Surajit}~\lnm{Mondal}}
\address[id=aff1]{National Centre for Radio Astrophysics, Tata Institute of Fundamental Research, Pune, India}

\runningauthor{Mondal, Surajit}
\runningtitle{EUV counterparts of quiet sun radio transients}

\begin{abstract}
Recently nonthermal radio transients from the quiet sun have been discovered and it has been hypothesised using rough calculations that they might be important for coronal heating. It is well realized that energy calculations using coherent emissions are often subject to poorly constrained parameters and hence have large uncertainties associated with them. However energy estimates using observations in the extreme ultraviolet (EUV) and soft X-ray bands are routinely done and the techniques are pretty well established. This work presents our first attempt to identify the EUV counterparts of these radio transients and then use the counterpart to estimate the energy deposited into the corona during the event. I show that the group of radio transients studied here was associated with an brightening observed in the extreme ultraviolet waveband and was produced due to a flare of energy $\sim 10^{25}$ ergs. The fact that the flux density of the radio transient is only $\sim 2\,$mSFU suggests that it might be possible to do large statistical studies in future for understanding the relationship between these radio transients and other EUV and X-ray counterparts and also for understanding their importance in coronal heating.
\end{abstract}
\end{opening}

\section{Introduction} \label{sec:introduction}

Solar radio transients in general and particularly in the metric wavelengths, have been studied since the dawn of solar radio observations. However, to the best of my knowledge, these studies were mainly limited to radio bursts and highly energetic phenomenon like coronal mass ejections. These phenomenon are generally associated with active regions. For a review about the various types of solar radio bursts and their observational characteristics I refer the reader to \citet{wild1963}. These radio bursts are generally related to magnetic reconnection activities in the solar corona and hence at least some of them, especially the impulsive type I radio bursts, might have some connection to the hypothesised ``nanoflares", proposed to explain the coronal heating problem \citep[e.g.][etc.]{mercier1997, ramesh2013}. However, if the nanoflare hypothesis is true, then such transient emissions should also been present in the quiet sun. \citet{mondal2020}, henceforth referred to a M20, reported the first detection of such radio transients in the quiet sun. The reported radio transients were highly impulsive, narrowband and had flux densities generally in the range of a few tens of mSFU, several orders of magnitude weaker than the well studied solar radio bursts. Henceforth we shall refer to these emissions as Weak Impulsive Narrowband Quiet Sun Emission(s) or WINQSEs in short. Transient emissions from the quiet sun have also been reported in the EUV and X-ray band \citep[e.g.][etc.]{harrison1997, berghmans1998, krucker1998, benz2002, kuhar2018, chitta2021} and it will be very interesting to understand their relationship with WINQSEs.

Finding the counterpart of WINQSEs in the EUV and X-ray is also important from the perspective of coronal heating as well. M20 hypothesised that WINQSEs are coherent radio emission and are weak cousins of the well known Type III radio bursts. Like the type III radio bursts, the reconnection processes responsible for the WINQSEs deposit the largest fraction of energy into the corona and a minuscule fraction of energy is observed as coherent radio emission. Parameters required to estimate the deposited energy from the observed coherent radio emission suffers from large uncertainties \citep{subramanian2004}. Hence, it is very challenging to estimate the deposited energy from the radio emission alone. However, if we can identify the counterparts of WINQSEs in the EUV or X-ray data, we can use standard techniques to estimate the deposited energy, which is the key quantity of interest from a coronal heating perspective. 

In this paper I do a pilot study to identify the EUV counterpart of a group of WINQSEs and then use it to estimate the energy associated with the WINQSE group. This WINQSE group is a very tiny fraction of all the WINQSEs discovered in M20 and its counterpart was discovered in a serendipitous manner.

\section{Observations and data analysis}

The radio data used in this work has already been presented in M20. I refer the reader to M20 for details about the radio observations and data analysis procedure. Here, I will only give a brief summary of the details provided there.

\begin{figure}
    \centering
    \includegraphics[trim={2cm 3cm 1cm 3cm},clip,scale=0.85]{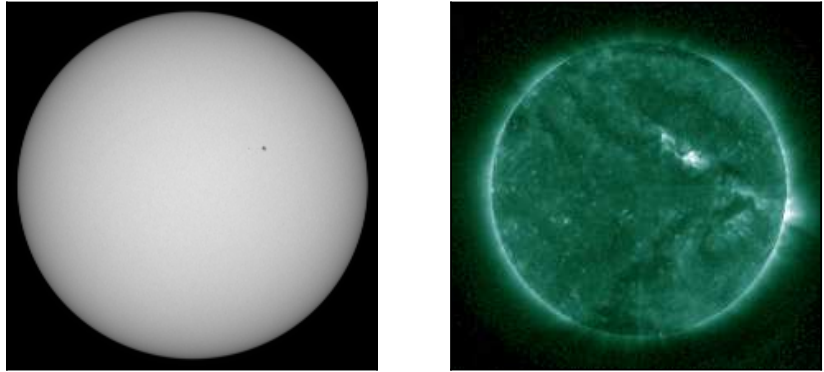}
    \caption{{Left panel: HMI magnetogram Right panel: Solar image obtained with the AIA at 94 \AA$\,$. Both data were taken near 01:40 UT on the day of the observation.}}
    \label{fig:AIA_images}
\end{figure}

These observations were taken on November 27, 2017. The sun was exceptionally quiet on this day \footnote{\url{https://www.solarmonitor.org/?date=20171127}}. In Fig. \ref{fig:AIA_images}, I have shown the magnetogram and solar image at 94 \AA$\,$. As is evident from Fig. \ref{fig:AIA_images}, Only one active region (NOAAA 12689) was present on the visible part of the solar disc on this day. M20 analysed 70 minutes of data at 4 frequencies near 96, 120, 132 and 160 MHz using the Automated Imaging Routine for Compact Array for Radio Sun \citep[AIRCARS,][]{mondal2019}. The images were made at every 0.5s and had a temporal resolution of 0.5s, spectral resolution of 160 kHz. In this paper, I use data corresponding to 132 MHz. The spatial resolution of the images at this frequency is $\sim 2.5^{'}$. In the full 70 minutes of data M20 detected $\sim 25000$ WINQSEs in this frequency. However, here I have focussed my attention on identifying the EUV counterpart of a group on WINQSEs visible near 01:58:14 UT and originating at the same location in the radio image. Henceforth for simplicity I will use WINQSEs and a group of WINQSEs interchangebly.


\section{Results}

\begin{figure}
\centering
\includegraphics[trim={7cm 2cm 0 0.5cm},scale=0.45]{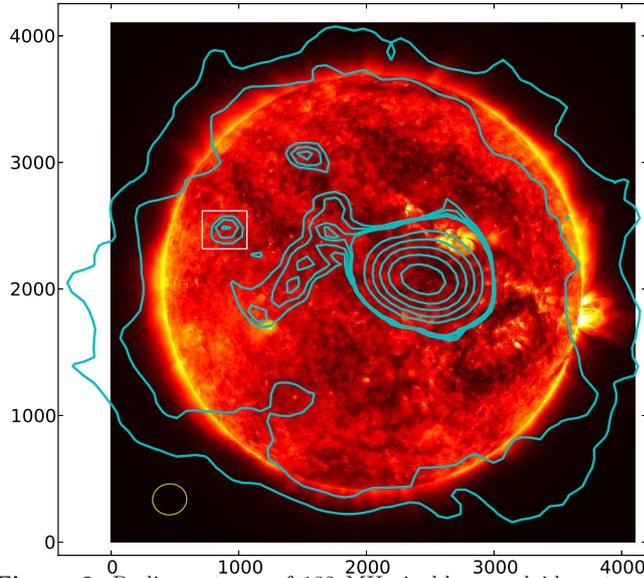}
\caption{Radio contours of 132 MHz in blue overlaid on top of a 171\AA $\,$image. The radio image is from 01:58:14.5 UT and has been generated using data spanning 0.5s and 160 kHz. The EUV map is from 01:58:11 UT. The yellow ellipse at the lower left corner shows the resolution of the radio map. The white boxes shows the two WINQSEs studied here.}
\label{fig:euv_radio_full_sun}
\end{figure}

\begin{figure*}
\centering
\includegraphics[trim={0 0 0.5cm 0},clip,scale=0.3]{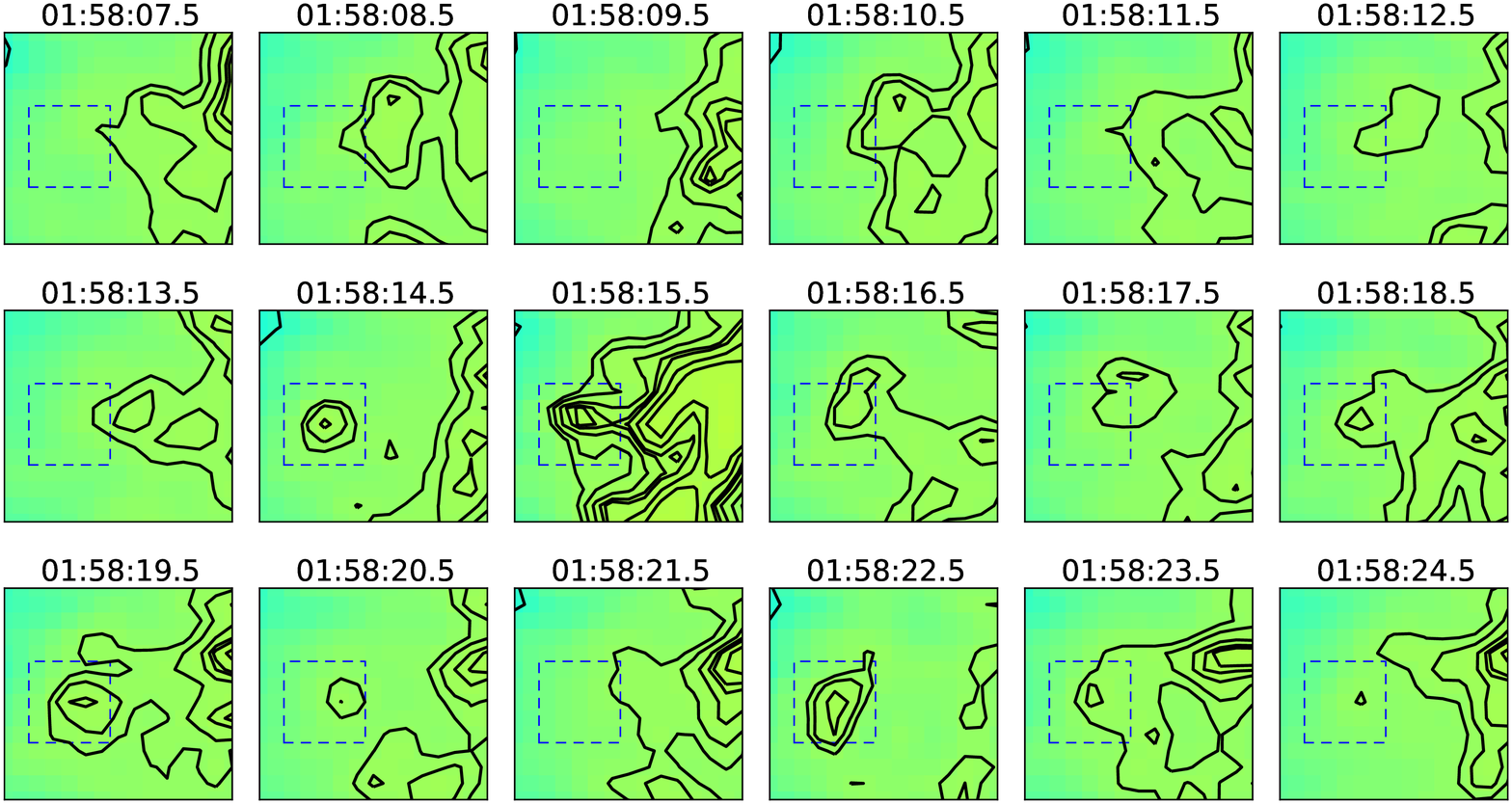}
\caption{Zoomed in view at the location of the WINQSE as a function of time. The dashed box shows the location of the WINQSE. The contours are radio intensity contours at 132 MHz. The contour levels are same in all the images shown. The times corresponding to each panel is give at the top of the corresponding panel.Each image is made using 0.5s data.}
\label{fig:winqse1_timeseries}
\end{figure*}

In Fig. \ref{fig:euv_radio_full_sun}, I have overlaid the radio contours at 01:58:14.5 UT in blue on top of a AIA image observed with the 171\AA $\,$ filter. All images are made using a data span of 0.5s. The radio compact sources enclosed by the white boxes is the location of the WINQSEs which will be studied here. The large radio source seen near the centre of the solar disc is a Type I noise storm associated with the active region NOAAA 12689. In Fig. \ref{fig:winqse1_timeseries}, I have shown how the flux at the location of the WINQSEs changed as a function of time. The dashed box shows the location of the WINQSEs. The contour levels are same in all the images shown. Multiple brightenings are visible inside the blue box, suggesting the occurrence of several WINQSEs in this time span. 

 We believe that WINQSEs are the radio counterparts of proposed nanoflares. Since nanoflares are expected to happen at very small spatial scales, we also expect WINQSEs to have a compact spatial morphology. Additionally, the quiet sun will always have a thermal contribution, and the nonthermal emission from the WINQSEs will always be observed on top of this thermal background. It is expected that the smooth quiet sun thermal background will have a large scale spatial morphology. Hence, as a conservative criterion, I assume that only compact structures are related to WINQSEs. Quantitatively, I classify a structure as compact if the following conditions are satisfied
 \begin{enumerate}
 \item The total area enclosed by all the closed contours surrounding a local peak is much smaller than the area of the solar disc.
 \item There is no second local peak within the closed contours.
 \end{enumerate}

\begin{figure}
    \centering
    \includegraphics[scale=0.35]{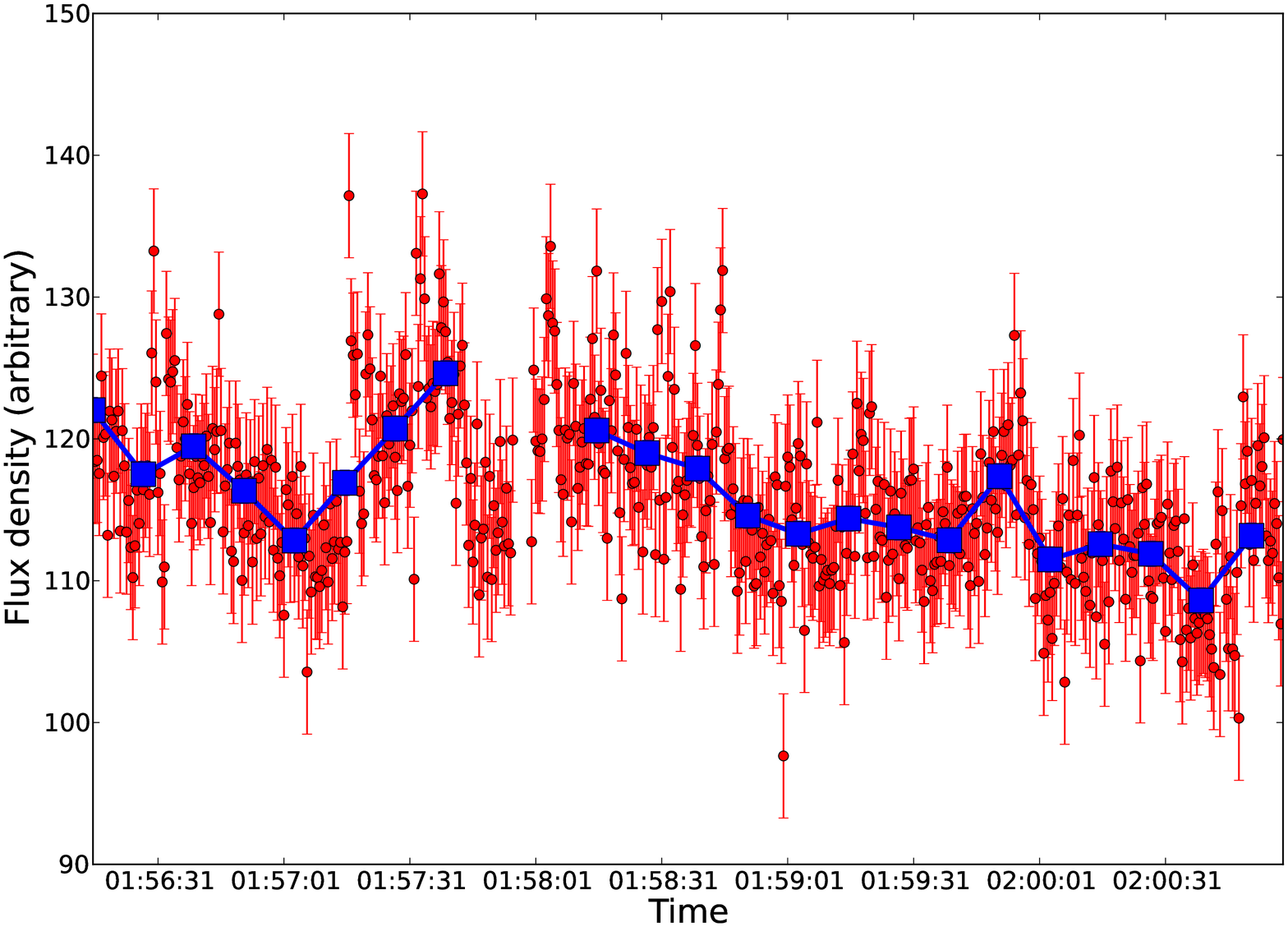}
    \caption{Red points show the observed/raw flux density variation at the location of the WINQSE at 0.5s resolution. The blue squares show the 12s smoothed light curve corresponding to the red points.}
    \label{fig:winqse1_quiet_sun_timeseries}
\end{figure}

It is evident from Fig. \ref{fig:winqse1_timeseries} that WINQSE like behaviour is not observed at times between 01:58:7.5--01:58:13.5. At 01:58:14 and 01:58:14.5 there is clear evidence of a WINQSE inside the blue dashed box. The structure reaches its peak flux density at 01:58:15.5 and then starts to decay. It starts to brighten again at 01:58:22, was brightest at 01:58:22.5 and can be regarded as undetected at 01:58:24.5. It is worth mentioning that M20 had demonstrated that WINQSEs are highly impulsive in nature and a single timeseries shows a large number of these structures in general. Our image plane analysis confirms this conclusion independently. The flux density timeseries at the location of the WINQSEs is shown in Fig. \ref{fig:winqse1_quiet_sun_timeseries}. The red points show the raw flux density, whereas the blue squares shows a 12s smoothed lightcurve.


Following M20, I have estimated the peak flux density of the WINQSEs to be $\sim 20$ Jy (1 Jy=$10^{-4}\,$SFU). I also find in this timerange there were 7 time instants when the calculated $\Delta F/F$ exceeded 0.1, which is the limiting value as found in M20. For these events $\Delta F/F$ ranged from 0.1--0.16. It should be noted that although the flux density of the WINQSE is as low as $20$ Jy, it has been detected at a significance level exceeding $4\sigma$, where $\sigma$ is the error in the flux density estimation and has been estimated from the image.

\begin{figure}
    \centering
    \includegraphics[scale=0.3]{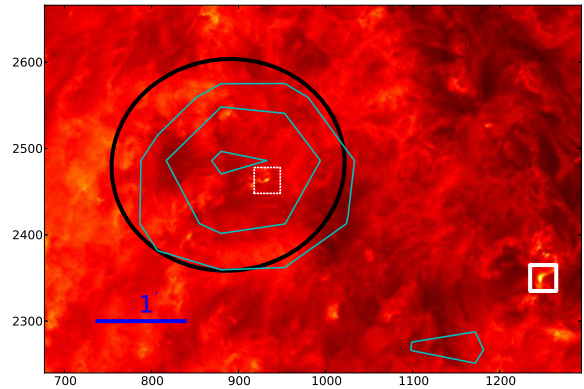}
    \caption{Radio contours of the WINQSE at 01:58:14.5 UT overlaid on AIA 171\AA $\,$ map. The black ellipse shows the angular resolution of the radio data. The EUV bright sources in the two white rectangles shows the two potential counterparts of the WINQSE. { The blue line shows the $1^{'}$ length scale.}}
    \label{fig:showing_both_sources}
\end{figure}

In Fig. \ref{fig:showing_both_sources}, I have shown the radio contours overlaid on a AIA 171\AA$\,$ map after zooming into the location of the WINQSEs. The black ellipse shows the angular resolution of the radio data. The EUV bright sources inside the two white boxes show variability in times similar to the WINQSEs and has been identified by visual inspection of the EUV data. I will refer to the EUV source inside the radio contour and bounded by a dashed rectangle as SOURCE1, and the EUV source inside the solid rectangle as SOURCE2. I choose to treat SOURCE2 as a potential counterpart of the WINQSE as it is possible that due to scattering or due to complex magnetic field connectivity \citep{kontar2019,mohan2019a,mohan2019b}, the observed position of the radio source has a significant spatial separation from its EUV counterpart. In subsequent paragraphs I have shown various properties related to these two sources, which I will use to try to identify the EUV counterpart of the WINQSEs. 

\begin{figure}
\centering
\includegraphics[scale=0.25]{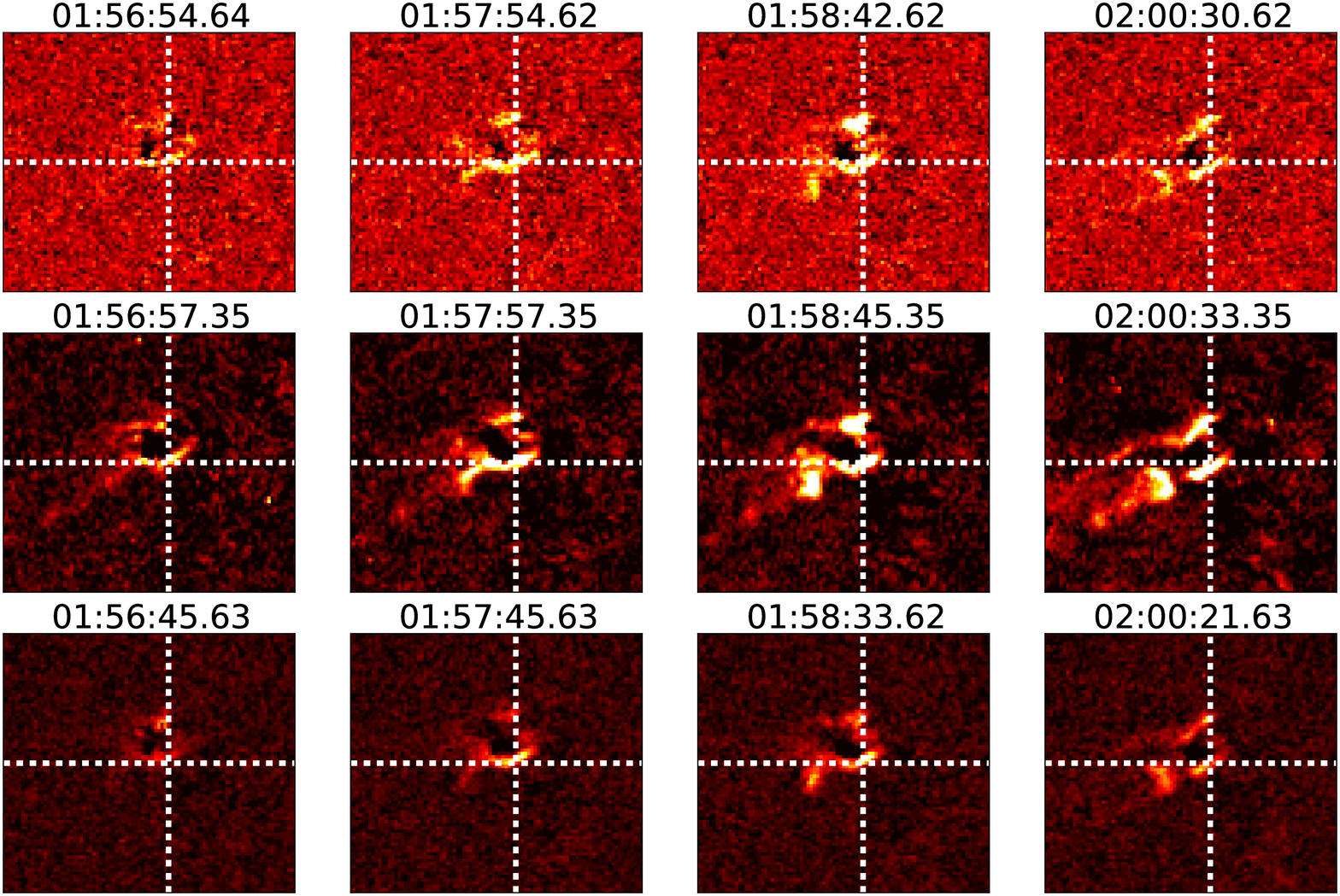}
\caption{Base difference images corresponding to SOURCE1. Upper panel: 131 \AA$\,$. Middle panel: 171 \AA$\,$. Lower panel: 211 \AA$\,$. The time of the AIA image is written on top of every panel. Image nearest to 01:55:30 has been subtracted for every filter. {The intersection point of the dashed white lines marks the region which has been used for all subsequent EUV analysis for SOURCE2, unless stated otherwise.}}
\label{fig:base_diff_euv_radio_overlay_source1}
\end{figure}

\begin{figure}
\centering
\includegraphics[scale=0.25]{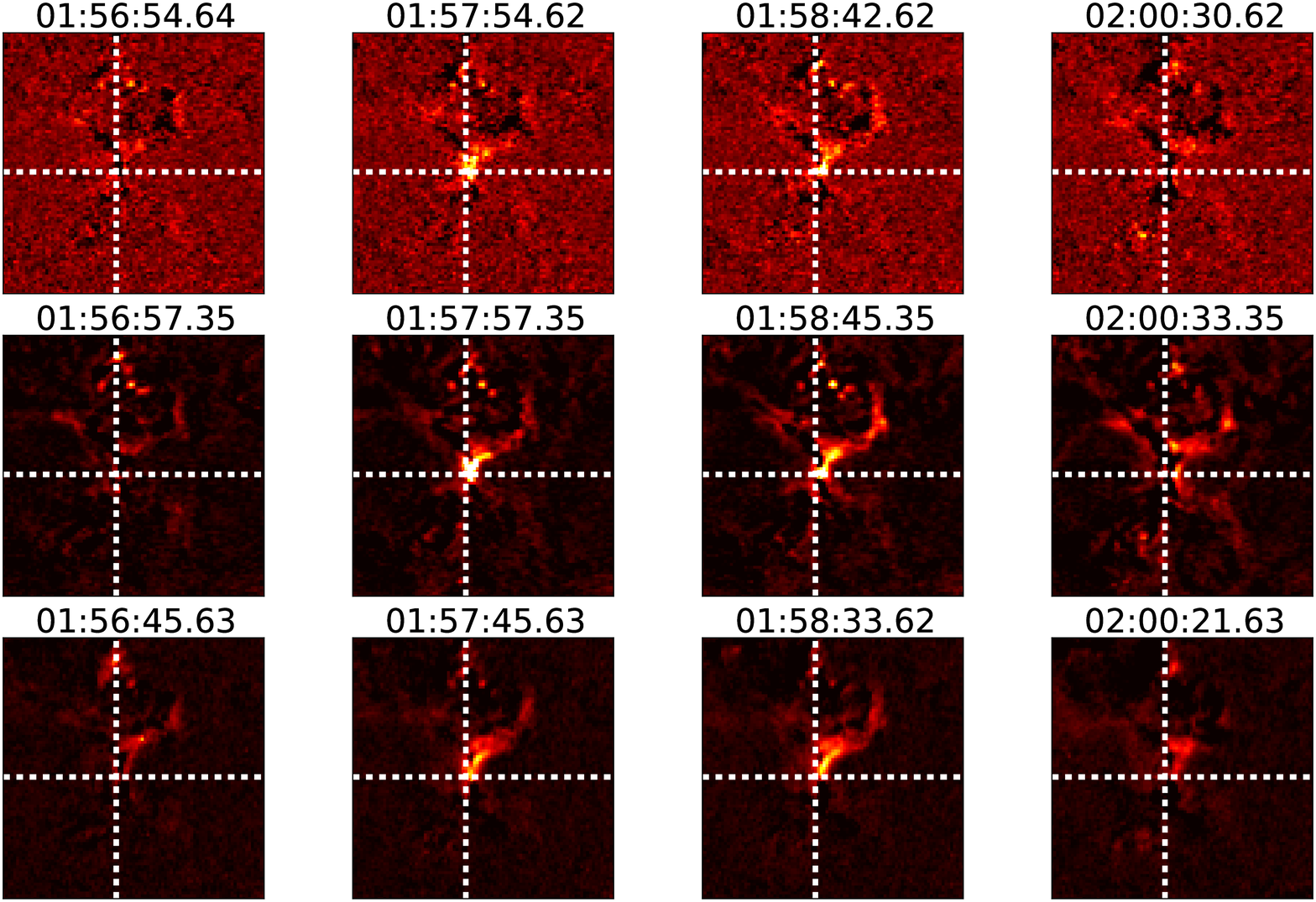}
\caption{Base difference images corresponding to SOURCE2. Upper panel: 131 \AA$\,$. Middle panel: 171 \AA$\,$. Lower panel: 211 \AA$\,$. The time of the AIA image is written on top of every panel. Image nearest to 01:55:30 has been subtracted for every filter. {The intersection point of the dashed white lines marks the region which has been used for all subsequent EUV analysis for SOURCE2, unless stated otherwise.}}
\label{fig:base_diff_euv_radio_overlay_source2}
\end{figure}

\begin{figure}
\centering
\includegraphics[scale=0.35]{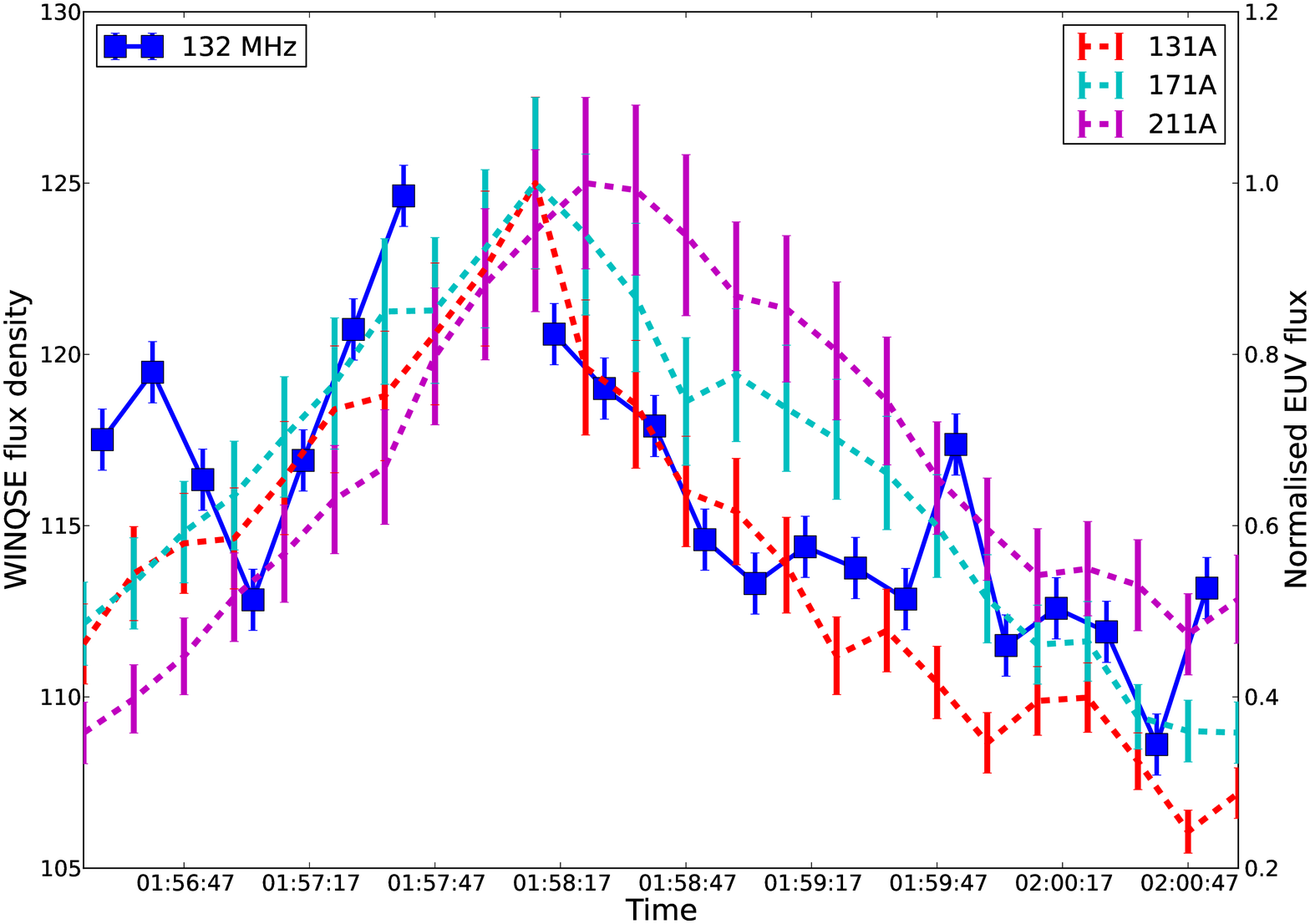}
\caption{The red, cyan and magenta dashed lines shows the normalised EUV flux variation of SOURCE1 at 131, 171 and 211 \AA$\,$ respectively. The blue line shows the WINQSE radio light curve smoothed over 12 seconds. }
\label{fig:EUV_radio_light_curve_source1}
\end{figure}

\begin{figure}
\centering
\includegraphics[scale=0.35]{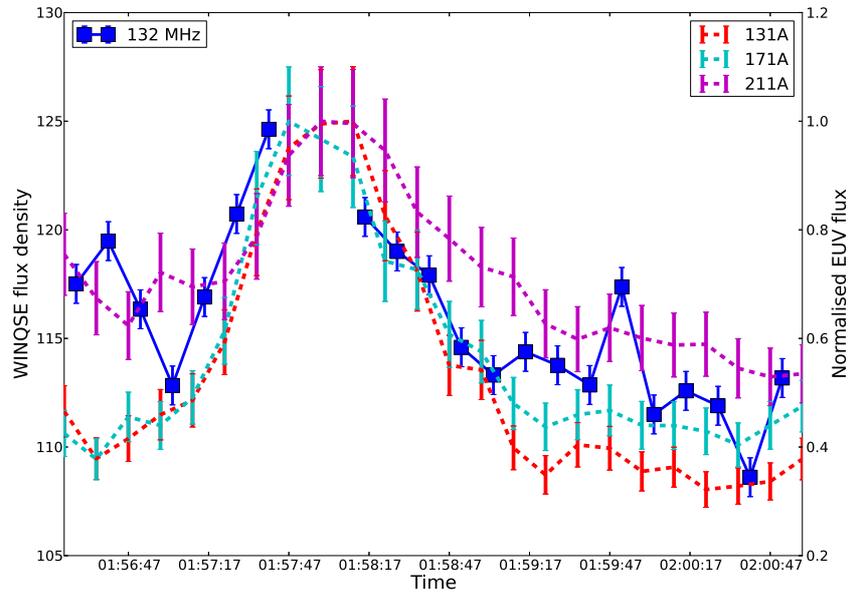}
\caption{The red, cyan and magenta dashed lines shows the normalised EUV flux variation of SOURCE2 at 131, 171 and 211 \AA$\,$ respectively. The blue line shows the WINQSE radio light curve smoothed over 12 seconds. }
\label{fig:EUV_radio_light_curve_source2}
\end{figure}

In Fig. \ref{fig:base_diff_euv_radio_overlay_source1} and \ref{fig:base_diff_euv_radio_overlay_source2}, I have shown base difference maps at 171, 131 and 211 \AA $\,$ after zooming into SOURCE1 and SOURCE2 respectively. The time of the base difference map is shown on top of each panel. The map closest to 01:55:30 UT has been subtracted to obtain the base difference map. { The intersection point of the two white dashed lines} shows the location of our interest in the EUV map. I show later that a box centred on this green star shows signature of energy enhancement at similar times as when we observe the radio enhancement. 
In Fig. \ref{fig:EUV_radio_light_curve_source1} and \ref{fig:EUV_radio_light_curve_source2}, I have plotted the normalized EUV flux density at 171 \AA, 131 and 211 \AA $\,$ within a box centred on the {point of interest} shown in Fig. \ref{fig:base_diff_euv_radio_overlay_source1} and \ref{fig:base_diff_euv_radio_overlay_source2} respectively. The errorbars in the EUV lightcurves have been obtained following \citet{boerner2012} and adding a 10\% systematic uncertainty in quadrature. The 12s smoothed corrected radio light curve at location of the WINQSE is shown using black points. 


\begin{figure}
     \centering
     \includegraphics[scale=0.45]{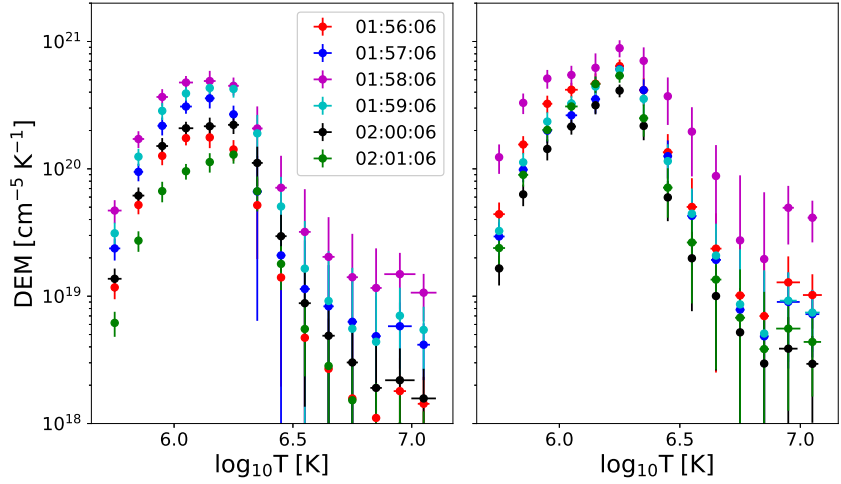}
     \caption{DEM distribution as a function of temperature for SOURCE1 (left panel) and SOURCE2 (right panel). }
     \label{fig:dem_time_variation}
 \end{figure}

I have performed a differential emission measure (DEM) analysis on these two sources to understand how the energy content at these two locations vary as a function of time. For this purpose, I use the publicly available code available from \url{https://github.com/ianan/demreg} based on \citet{hannah2012} and \citet{hannah2013}. The code does a regularised inversion of the data and returns the differential emission maps corresponding to each pixel. I averaged the data over the region of interest and have shown the DEM curves at different times corresponding to SOURCE1 and SOURCE2 in Fig. \ref{fig:dem_time_variation}. From these figures it is evident that the DEM for both these sources peaks around 01:58:00 UT.

Using DEM it is possible to estimate the energy deposited into the corona and is given by,
 
 \begin{eqnarray}
     E&=& 3<n_e>k_B<T>L^3 \\
     <n_e>&=& \sqrt{EM/L} \\
     EM&= & \int DEM(T) dT \\
     <T>&=& \frac{\int DEM(T)T dT}{EM}
 \end{eqnarray}
 
\begin{figure}
     \centering
     \includegraphics[scale=0.45]{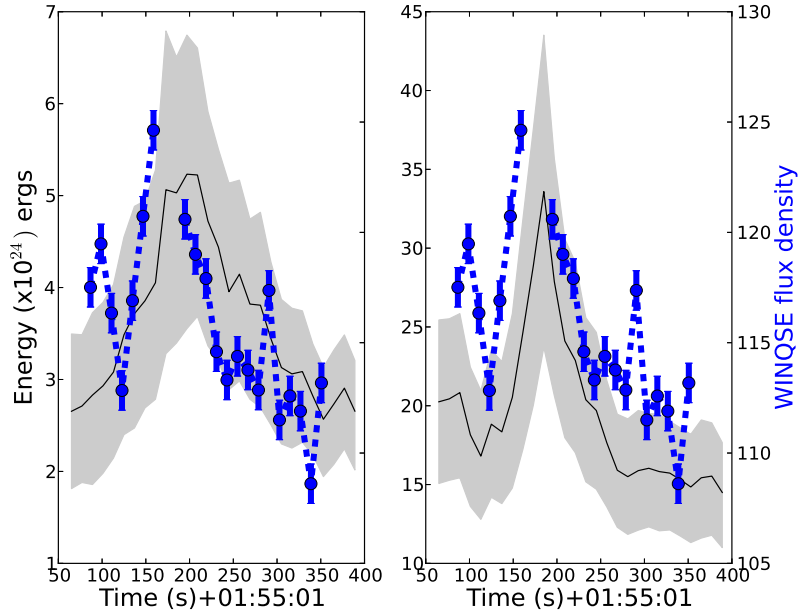}
     \caption{Shows the energy variation in SOURCE1 (left panel) and SOURCE2 (right panel). The blue curve shows the 12s smoothed lightcurve at the location of the WINQSE. }
     \label{fig:energy_variation_both_sources}
 \end{figure}

 Here L is the length of the emitting region. It is clear that while DEM is obtained by modelling the EUV light curve at various filters, the emitting length scale has to be obtained from the EUV images. I find that emitting length for SOURCE1 and SOURCE2 is 1.3 Mm and 2.3 Mm respectively. Using these numbers I have estimated the energy for both the sources and have plotted them in Fig. \ref{fig:energy_variation_both_sources}. The grey region shows the variation of the energy due to the random and systematic uncertainties in the AIA data. The two bounding curves of this error region has been computed by using the $DEM+\Delta DEM$ and $DEM-\Delta DEM$ for all filters, where $\Delta DEM$ is the error on DEM returned by the code.
 
 \section{Discussion}
 
 \subsection{Finding the counterpart of the WINQSE}
 
 It is evident from Figs \ref{fig:EUV_radio_light_curve_source1}, \ref{fig:EUV_radio_light_curve_source2} and \ref{fig:energy_variation_both_sources} that both SOURCE1 and SOURCE2 show temporal variation very similar to the smoothed radio lightcurve of the WINQSE group. This implies that both of these sources are possible counterparts of the WINQSE group. Additionally there is another possibility that the two sources are connected by a coronal loop and then it is might be possible to associate both the sources to the WINQSE group. I have discussed each of these scenarios and provided their pros and cons in the following subsections.
 
 \subsubsection{Scenario 1: SOURCE1 and SOURCE2 are connected}
 
\begin{figure}
     \centering
     \includegraphics[scale=0.5]{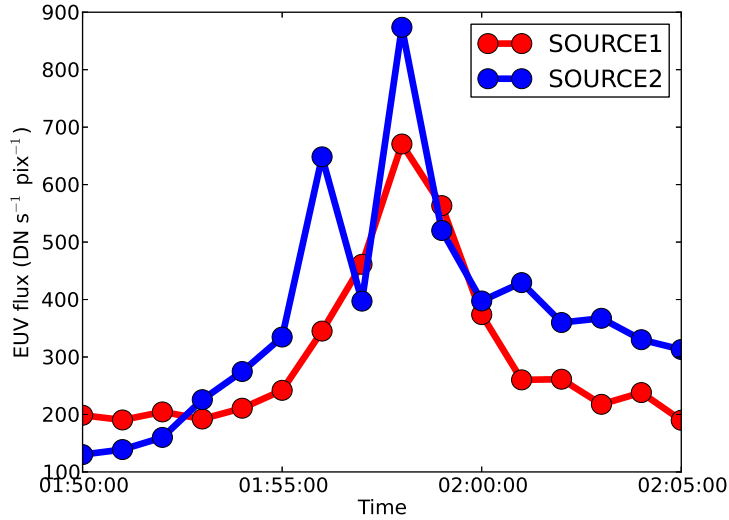}
     \caption{Light curve of SOURCE1 and SOURCE2 at 171\AA $\,$.}
     \label{fig:long_term_evolution}
 \end{figure}

 If both of the EUV sources are connected by some coronal loop, then it can very easily explain why they show variability at similar timescales. Since coronal loops are quite long lived, it is expected that this correlation will also hold for some significant amount of time. To investigate if this prediction is true or not, in Fig. \ref{fig:long_term_evolution} I show the EUV light curve of both these sources. From the figure it is evident that while the peak near 01:58:00 UT is present at both the sources, SOURCE2 shows an additional peak near 01:56, which is absent in SOURCE1. This observation suggests that there is a high probability that SOURCE1 and SOURCE2 are not connected with each other and it is mere chance that the two peaks line up.
 
 \subsubsection{Scenario 2: Either SOURCE1/SOURCE2 is the counterpart of WINQSE1}
 
\begin{figure}
     \centering
     \includegraphics[scale=0.4]{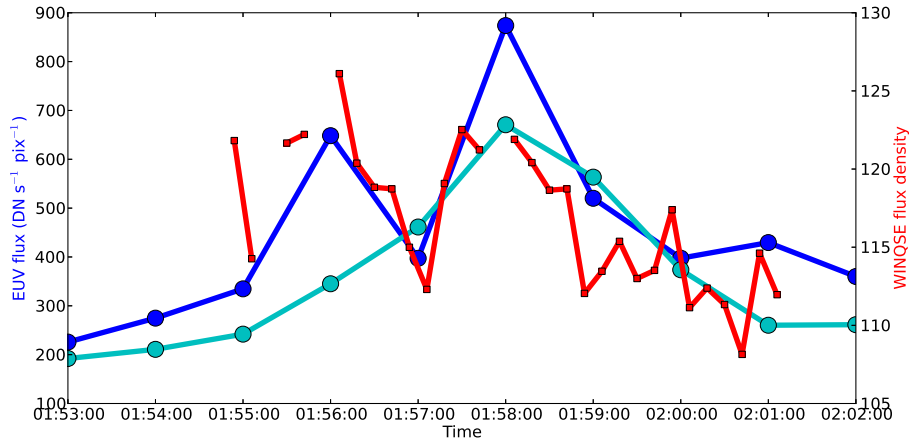}
     \caption{Light curve of SOURCE1 (cyan curve), SOURCE2 (blue curve) and the WINQSE group (red curve). }
     \label{fig:aia_radio_long_term_lightcurve}
 \end{figure}

 It is evident from Fig. \ref{fig:showing_both_sources}, while SOURCE1 is co-located with WINQSE1 in the sky plane, SOURCE2 is separated by $\sim$3--4 arcminutes. However, shifts of similar magnitudes have already been observed in previous studies of Type III radio bursts and since scattering only depends on the intervening medium, this can also affect WINQSEs. Additionally, it is expected that the radio emission due to the WINQSE group is originating from a much larger height ($\sim 1.2R_\odot$), compared to the EUV sources. Hence depending on the exact orientation of the magnetic field loop connecting the WINQSE location and its counterpart, the angular separation between the WINQSE and its counterpart can also vary. Due to these reasons, I believe that spatial co-location is not a strong discriminator in determining the counterparts, if their separation from WINQSE is a few arcminutes. Hence, I will try to use the temporal and energy properties of both sources to determine which one of these is the counterpart. However, here also there is a roadblock. Both SOURCE1 and SOURCE2 show variation in energy at similar timescales to the variability in the smoothed WINQSE light curve. Hence based on the temporal variability it seems that these two sources are equally likely to be counterpart of WINQSE. However, it must be noted that even at the location of the WINQSE group, multiple WINQSEs happened during the full 70 minutes of data. To break the degeneracy between the two EUV sources which show variation at same timescales as the group of WINQSEs between 01:56:30- 02:00:00 UT, I have used data slightly before 01:56:30 UT. It is clear from Fig. \ref{fig:long_term_evolution} that whereas SOURCE2 shows a small brightening around 01:56 UT, SOURCE1 does not show any such signature. It is not expected that the coronal loop connectivity or scattering properties will change significantly in this small time separation of about 2--3 minutes. Hence if SOURCE2 is the counterpart then I expect to see a similar enhancement in the WINQSE smoothed radio lightcurve as well around 01:56 UT. On the contrary if this expectation is not met, then it will be more likely that the SOURCE1 is the counterpart of the WINQSE. In Fig. \ref{fig:aia_radio_long_term_lightcurve} I have shown the radio light curve covering a much larger timerange than shown earlier. Multiple images between 01:55--01:57 UT had dynamic range limitations and hence are not shown in the light curve. In spite of this, there is strong indication that the smoothed radio light curve shows enhancement near 01:56 UT. Based on this evidence I conclude that SOURCE2 is the EUV counterpart of the WINQSE group.
 
 \subsection{Physical model of WINQSEs}
 
 M20 hypothesised that WINQSEs are the weaker cousins of the emissions observed by \citet{mohan2019b}. Here I present strong evidence, albeit in a two groups of WINQSEs, that the smoothed radio light curve shows variation at very similar timescales as its EUV counterpart. The energy variation in the system also follows a similar trend. This is exactly what was observed by \citet{mohan2019b}. Based on this additional evidence, I present a physical model below.
 
 \begin{figure}
      \centering
      \includegraphics[trim={0 5cm 0 0},clip,scale=0.5]{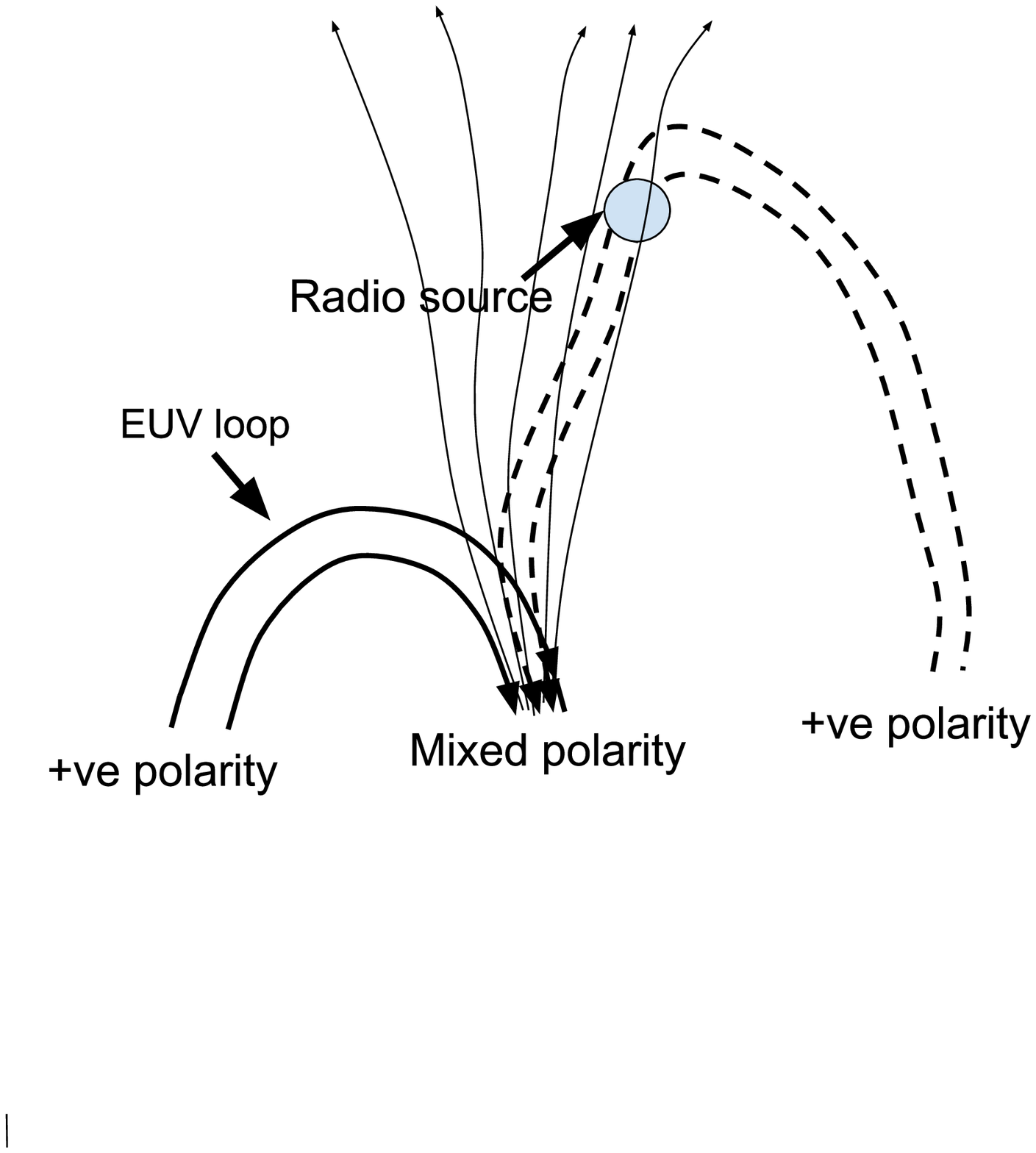}
      \caption{Cartoon drawing showing the location of the radio source and the bright loop observed in the EUV.}
      \label{fig:flux_tube_geometry}
  \end{figure}
 
As shown in Fig. \ref{fig:flux_tube_geometry}, I hypothesise that the EUV bright loop and the loop containing the radio source has at least a common footpoint. It should be noted that I have added oppositely directed open field linear near this footpoint, to suggest that this footpoint has a complex polarity, with presence of both +ve and -ve regions. Due to this reconnections will happen at the footpoint and this will heat the EUV loop and thus it will become bright in the EUV. These reconnections will also induce magnetic waves which will then travel along the magnetic tube and induce reconnections at different heights. These reconnections generates additional nonthermal electron beams which will then emit in the radio band. Since the original reconnections at the magnetic footpoint are weak, then the reconnections they induce higher up in the atmosphere close to the radio source are also weak. Hence the electron beams generated are weak and gets easily damped due to collisions, explaining the narrow bandwidths of the WINQSEs. This model also shows why the radio source and the EUV source can have intrinsic separation in the sky plane. This model is similar to the one proposed my M19 and I refer the reader to M19 and \citet{mohan2021} for more details. It is noted that the presence of this mixed polarity region is not the only way to induce reconnections in the EUV loop. For example, reconnections can also happen due to strong twist in the field lines in the EUV loop. Hence as long as reconnections happen near the footpoint of the EUV loop, this model will work. A key prediction of this model is that the radio and X-ray data will show some lag, which is equal to the time required for the magnetic wave to travel to the height of the radio source from the magnetic footpoint. This prediction will be tested in future using simultaneous X-ray, EUV and radio data. 

\subsection{Possible relationship with campfires}

Recently the Solar Orbiter has observed transient brightenings in the quiet sun \citep{berghmans2021} and have termed them as ``campfires". \citet{berghmans2021} found that these brightenings are also visible in AIA bands like 193, 171, 211 and 304 \AA $\,$ and are generally not observed in the other bands. However there are also suggestions that at least some campfires also show signatures in other bands as well \citep{rutten2020}. Campfires have a linear extension of 0.2--4 Mm. The EUV counterpart studied here has a linear extension 2 Mm and lasts for $\sim$2--3 minutes, again in the range of campfires. The peak flux observed in 171 \AA $\,$ is $\sim 900$DN/s. Converting this to the peak photon count using a conversion factor of 1.168 \citep{boerner2012}, I obtain that at the peak, 1051 photons were detected each second. This lies near the peak of the probability of total intensity distribution of campfires \citep{berghmans2021}. Based on these evidences I conclude that it is highly likely that the observed EUV counterpart is also a "campfire" and hence in corollary, WINQSEs might also be associated with "campfires". Recent simulations also hint towards a relationship between the two phenomena. \citet{chen2021} found three magnetic field orientations which can produce campfires and one of them match almost exactly with the emission geometry found in \citet{mohan2019b} and \citet{mohan2021}. Future studies with simultaneous observations with the Solar Orbiter and sensitive radio interferometers will be able to investigate possible relationships between these two phenomena in detail. 

\section{Conclusion}

In this article, I have identified the EUV counterpart of a group of WINQSEs and estimated that during the process, $\sim 10^{25}$ ergs were deposited into the corona. While the relationship between EUV flares and radio bursts are well studied, this is the first study to understand the EUV counterparts of the WINQSEs. To the best of my knowledge, this is the weakest EUV transient event known to have a radio counterpart as well. However, the energy estimated for this event is about 1--2 orders of magnitude higher than the nanoflares studied in \citet{chitta2021} and is not exactly in the nanoflare regime. While the events studied here is on the less energetic side of the events reported in M20, further work have revealed the presence of much weaker features in the data than found in M20 and will be presented in an upcoming publication (Biswas et al., in preparation). Hence, this work in no way rules out the hypothesis put forth by M20 that WINQSEs are the radio counterparts of nanoflares. On the contrary this work goes on to show that it is possible to associate EUV counterparts to WINQSEs and thus encourages further work towards this direction.

\begin{acks}
 This scientific work makes use of the Murchison Radio-astronomy Observatory, operated by the Commonwealth Scientific and Industrial Research Organisation (CSIRO). I acknowledge the Wajarri Yamatji people as the traditional owners of the Observatory site. Support for the operation of the MWA is provided by the Australian Government through the National Collaborative Research Infrastructure
Strategy (NCRIS), under a contract to Curtin University administered by Astronomy Australia Limited.  I gratefully acknowledge Divya Oberoi (NCRA-TIFR) for coming up with the acronym WINQSE. I would also like to thank the referee for his/her insightful commnents which have helped me to make the manuscript immensely better. I acknowledge the Pawsey Supercomputing Centre, which is supported by the Western Australian and Australian Governments. I acknowledge support of the Department of Atomic Energy, Government of India, under the project no. 12-R\&D-TFR-5.02-0700.
The SDO is a National Aeronautics and Space Administration (NASA) spacecraft, and I acknowledge the AIA science team for providing open access to data and software. 
This research has also made use of NASA's Astrophysics Data System (ADS). 
I thank the developers of Python 2.7\footnote{See
https://docs.python.org/2/index.html.} and the various associated packages, especially Matplotlib\footnote{See http://matplotlib.org/.}, Astropy,\footnote{See http://docs.astropy.org/en/stable/.} and NumPy\footnote{See https://docs.scipy.org/doc/.}. 
\end{acks}

\bibliographystyle{spr-mp-sola}
\bibliography{bibliography}

\end{article} 

\end{document}